\documentclass[a4paper]{article}
\usepackage{multirow}

\usepackage{INTERSPEECH2021}

\title{The DKU-DukeECE-Lenovo System for the Diarization Task \\ of the 2021 VoxCeleb Speaker Recognition Challenge}
\name{Weiqing Wang$^{1, 3}$, Danwei Cai$^{1, 3}$, Qingjian Lin$^2$, Lin Yang$^2$, Junjie Wang$^2$, Jin Wang$^2$,  Ming Li$^1$}

\address{
  $^1$ Data Science Research Center, Duke Kunshan University, Kunshan, China \\
  $^2$ AI Lab, Lenovo Research, Beijing, China  \\
  $^3$ Department of Electrical and Computer Engineering, Duke University, Durham, USA}
\email{\{weiqing.wang, danwei.cai, ming.li369\}@duke.edu,\\ \{linqj3, yanglin13, wangjj9, wangjin19\}@lenovo.com}

\begin{document}

\maketitle
\begin{abstract}
This report describes the submission of the DKU-DukeECE-Lenovo team to the VoxCeleb Speaker Recognition Challenge (VoxSRC) 2021 track 4. Our system includes a voice activity detection (VAD) model, a speaker embedding model, two clustering-based speaker diarization systems with different similarity measurements, two different overlapped speech detection (OSD) models, and a target-speaker voice activity detection (TS-VAD) model. Our final submission, consisting of 5 independent systems, achieves a DER of 5.07\% on the challenge test set. 
\end{abstract}

\section{Introduction}

As the speaker embedding becomes more and more robust, the conventional diarization system also achieves good performance since the speaker confusion has been significantly reduced. To further improve the performance and reduce diarization error rate (DER), many researches focus on overlapped speech detection (OSD) to reduce the missed speaker error, including speech separation \cite{microsoft}, overlap detection \cite{overlap}, end-to-end neural speaker diarization (EEND) \cite{eend} and target-speaker voice activity detection (TS-VAD) \cite{tsvad}. 

We also explore many different OSD models in this challenge. First, a ResNet-based model with LSTM back-end is employed for overlap detection, where the output is 1 for overlapped speech and 0 otherwise. Second, an x-vector- and ResNet-based TS-VAD model is used to refine the output from the conventional diarization systems. Finally, we propose a 2-speaker TS-VAD model for overlap detection, where a pair of speaker embeddings are fed to the TS-VAD model, and the overlapped speech regions between these two speakers are detected. Compared with the original TS-VAD, this method is not restricted to the number of speakers.

\section{Dataset Description}

Since our TS-VAD model takes a long time for inference, we only use the last 46 recordings in the test dataset as our validation dataset, referred to as VAL46. The development dataset with the remaining data in the test dataset is used as our development dataset, which contains 402 recordings and is referred to as DEV402. The detailed dataset used in this challenge for each model include:

\begin{itemize}

\item Voice activity detection (VAD): AMI \cite{AMI}, ICSI \cite{ICSI}, ISL (LDC2004S05), NIST (LDC2004S09), SPINE1\&2 (LDC2000S87, LDC2000S96, LDC2001S04, LDC2001S06, LDC2001S08), AISHELL-4 \cite{aishell4}, DIHARD II \cite{DIHARDII} and DIHARD III \cite{DIHARDIII} are the mixed training set. DEV402 and VAL46 is used for fine-tuning and validation, respectively. 
\item Speaker embedding: We use Voxceleb 1 \& 2 \cite{Voxceleb} as the training set.
\item Agglomerative hierarchical clustering (AHC): We directly tune the parameters on DEV402. 
\item LSTM-based similarity measurement with spectral clustering: We use the same dataset as VAD does. 
\item Overlap detection: We use DEV402 for training and VAL46 for validation. 
\item 2-speaker TS-VAD \& TS-VAD: These models are first trained on the data simulated by Librispeech \cite{Librispeech}. Then we transfer the model to VoxConverse \cite{Voxconverse2020} dataset with the data simulated by DEV402. Finally, we fine-tune the model on DEV402 and validate it on VAL46. 
\item Data augmentation: We perform data augmentation with MUSAN \cite{MUSAN} and RIRs \cite{RIRs} corpus.  

\end{itemize}

\begin{figure*}[t]
  \centering
  \includegraphics[width=\linewidth]{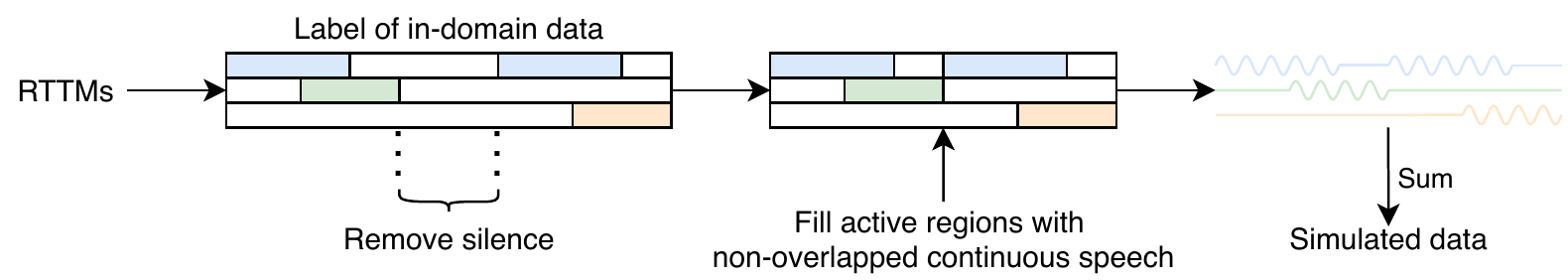}
  \caption{The data simulation strategy for TS-VAD.}
  \label{fig:simulation}
\end{figure*}
\section{Detailed Model Configuration}
The model architectures of VAD, overlap detection, and TS-VAD are very similar, including a ResNet34 \cite{resnet}, a statistical pooling layer, two BiLSTM \cite{bilstm} layers and two fully-connected layers with sigmoid function. For ResNet34, the channel width is \{32, 64, 128, 256\} with kernel size of 3. Each BiLSTM layer contains 256 units per direction with a dropout rate of 0.1. The two fully connected layers contain 128 and 1 unit, respectively.

\subsection{VAD}
We use a ResNet34 as the front-end model to extract the frame-level feature map. Next, a statistical pooling layer is employed on the feature map every $S$ frames. Finally, two BiLSTM layers and two fully-connected layers with a sigmoid function produce the posterior probability of speech. In our experiments, the input is 8s chunked wav, and the acoustic feature is 80-dim log Mel-filterbank energies with a frame length of 25ms and a frameshift of 10ms. 

The model is first trained on the mixed training set for 100 epochs with a learning rate of 0.0001 and then fine-tuned on DEV402 for 100 epochs with a learning rate of 0.00001 with binary cross-entropy (BCE) loss and Adam optimizer. We train four models with different $S=\{1, 2, 4, 8\}$ to obtain the VAD outputs with different resolutions. Finally, we average the outputs from these four models in frame level. The threshold for speech decision is set to 0.6. Table \ref{tab:vad} shows the false alarm (FA) and miss detection (MISS) on VAL46. 

\begin{table}[htb]
  \caption{The false alarm (FA), miss detection (MISS) and accuracy of the VAD model}
  \label{tab:vad}
  \centering
  \begin{tabular}[c]{lccc}
    \toprule
     & \textbf{FA [\%]} & \textbf{MISS [\%]} & \textbf{Accuracy} [\%] \\
    \midrule
    S=1    & 2.57 & 1.56 & 96.33 \\
    S=2    & 2.97 & 1.30 & 96.17 \\
    S=4    & 3.22 & 1.32 & 95.93 \\
    S=8    & 3.40 & 1.59 & 95.54 \\
    Fusion & 2.37 & 1.66 & 96.43 \\
    \bottomrule
  \end{tabular}
\end{table}

\subsection{Speaker Embedding}
The ResNet34 structure is employed as the front-end pattern extractor, which learns a frame-level representation from the input acoustic feature. A global statistic pooling (GSP) layer projects the variable length input to the fixed-length vector. Next, a 128-dim fully connected layer is adopted as the speaker embedding layer. The ArcFace \cite{arcface} (s=32,m=0.2) is used as a classifier. The detailed configuration of the neural network is the same as \cite{ffsvc20}. Table \ref{tab:eer} shows the EER on the Voxceleb1 test set. The input is 2$\sim$4s chunked wav, and the acoustic feature is 80-dim log Mel-filterbank energies with a frame length of 25ms and a frameshift of 10ms. 

\begin{table}[htb]
  \caption{The EER of the speaker embedding model}
  \label{tab:eer}
  \centering
  \begin{tabular}[c]{lc}
    \toprule
    \textbf{Training data} & \textbf{EER (\%)} \\
    \midrule
    Voxceleb 2 & 1.23 \\
    \bottomrule
  \end{tabular}
\end{table}

\subsection{Clustering-based System}
\subsubsection{AHC}
We use a similar AHC method as \cite{microsoft} without speech separation. First, speaker embeddings are extracted from the uniformly segmented speech with a length of 1.28s and shift of 0.32s, and two consecutive segments are merged into a longer segment if the distance is greater than a segment threshold. The pairwise similarity is measured by cosine distance. Next, we perform a plain AHC on the similarity matrix with a relatively high stop threshold to obtain the clusters with high confidence. These clusters are split into ``long clusters'' and ``short clusters'' by the total duration in each cluster, and the central embedding of each cluster is the mean of all speaker embeddings in the cluster. Later, each short cluster is assigned to the closest long cluster by the cosine distance of central embedding. Finally, if a short cluster is too different from all long clusters, which means that the distance between them is lower than a speaker threshold, we treat it as a new speaker. 

All of these parameters are directly tuned on the DEV402 by grid search. In our experiments, the segment threshold is 0.54, the stop threshold is 0.62, the duration for classifying long and short clusters is 6s, and the speaker threshold is 0.2. 

\subsubsection{LSTM-based Similarity Measurement with Spectral Clustering}
We use the same LSTM-based model as \cite{LinLSTM}. The model consists of two BiLSTM and two fully connected layers with a sigmoid function. Speaker embeddings are also extracted from the uniformly segmented speech with a length of 1.28s and shift of 0.64s. Next, the speaker embedding sequence $[\mathbf{x}_1, \mathbf{x}_2, ..., \mathbf{x}_n]$ is concatenated with repeated $\mathbf{x}_i$ as the input and produce the i-th row of the affinity matrix $\mathbf{S}$:

\begin{equation}
    \label{eq:BiLSTM}
    \mathbf{S}_i = [\mathbf{S}_{i,1}, \mathbf{S}_{i,2}, ... , \mathbf{S}_{i,n}] = f(
    \left[\begin{matrix}\mathbf{x}_i\\\mathbf{x}_1\end{matrix}\right], 
    \left[\begin{matrix}\mathbf{x}_i\\\mathbf{x}_2\end{matrix}\right], ...,
    \left[\begin{matrix}\mathbf{x}_i\\\mathbf{x}_n\end{matrix}\right]),
\end{equation}
where $f$ is the LSTM-based neural network, n is set to 64 in our experiments. More details can be found in \cite{DIHARDII-LSTM}. 

The model is trained on the mixed training set for 100 epochs and fine-tuned on DEV402 for 100 epochs. The model is optimized with BCE loss and Adam optimizer with a learning rate of 0.001. After obtaining the affinity matrix $\mathbf{S}$, we employ spectral clustering (SC) to get the final diarization results. 

\subsection{Overlap Detection}
The overlap detection model and training process are the same as that of the VAD model except for the training data, labels, and resolutions. We train the overlap detection model on DEV402, and we only average the outputs from two models with $S=\{1, 2\}$ since the overlapped regions are much shorter than the speech regions in the VAD task. The label is 1 for overlapped speech and 0 otherwise. After an overlapped region is detected, we replace the label with two closest speakers near this region. The threshold for overlap decision is set to 0.8. The input is 8s chunked wav, and the acoustic feature is 80-dim log Mel-filterbank energies with a frame length of 25ms and a frameshift of 10ms. 

\begin{table*}[tp]
  \caption{ The performance of different speaker diarization systems. $^*$Since the VAD model used in the 1st submissions is not well trained, the improvement of DER compared with other systems may come from both overlap detection (OD) and VAD on the test set. }

  \label{tab:result}
  \centering
  \begin{tabular}[c]{lcccccccc}
    \toprule
     \multirow{2}*{\textbf{Model}} & {\textbf{Submission}} &  \multicolumn{2}{c}{\textbf{VAL46}} & \multicolumn{2}{c}{\textbf{VoxSRC21 test set}}\\
      \cmidrule(lr){3-4} \cmidrule(lr){5-6} \cmidrule(lr){7-8} 
     
      & & \textbf{DER[\%]}  & \textbf{JER[\%]} & \textbf{DER[\%]}  & \textbf{JER[\%]} &\\
 
   \midrule
   Baseline                        & - &   -  &  -    & 17.99 & 38.72 \\
   \midrule
   AHC + Oracle VAD                & - & 2.61 & 21.93 & - & - \\
   LSTM-SC + Oracle VAD            & - & 3.16 & 28.04 & - & - \\
   \midrule
	1 AHC                          & 1$^*$ & 4.42 & 26.42 & 6.77$^*$ & 27.66$^*$  \\
	
	2 LSTM-SC                      & - & 4.97 & 31.04 & - & -  \\
	
	3 AHC + OD                     & - & 3.98 & 26.08 & - & - \\
	
	4 LSTM-SC + OD                 & - & 4.58 & 30.70 & - & - \\
	
	5 AHC + 2-spk TS-VAD as OD     & 4 & 3.96 & 25.82 & 5.45 & 27.55  \\
	
	6 LSTM-SC + 2-spk TS-VAD as OD & - & 4.56 & 30.51 & - & -  \\
	
	7 System 3 + TS-VAD            & - & 4.53 & 28.39 & - & -  \\
	
	8 System 5 + TS-VAD            & - & 4.51 & 28.33 & - & -  \\
	 \midrule
	 Fusion (3+4+7) & 2 & 3.93 & 25.68 & 5.36 & 29.10  \\
	 Fusion (3+4+5+7) & 3 & 4.02 & 27.11 & 5.82 & 29.78  \\
	 Fusion (3+4+5+6+8) & 5 & 4.10 & 30.97 & 5.07 & 29.16  \\
	  
     \bottomrule
     \end{tabular}
\end{table*}

\subsection{TS-VAD}
\label{sec:TSVAD}

\subsubsection{Data Simulation}
We simulate two 900-hour pre-training datasets. One is simulated from LibriSpeech, and another is simulated from DEV402. To obtain a simulated dataset that is similar to the VoxConverse dataset, we first extract the label from the DEV402, which is a matrix of size $T \times N$, where $T$ is the number of frames and $N$ is the number of speakers. Then, for the label of each speaker, we fill the active regions with a single speaker's speech from LibriSpeech or DEV402. Finally, we take the sum of all speeches as one simulated recording. Note that the non-speech regions in the labels are first removed. Figure \ref{fig:simulation} shows an example of the process of data simulation. 

\subsubsection{Training details}
TS-VAD has achieved an excellent performance on CHIME6 \cite{tsvad} and DIHARD III \cite{dihard3tsvad} challenge. Unlike the previous method using i-vector, we employ ResNet-based x-vector as the target-speaker embedding. 

The TS-VAD model is also similar to the VAD model except that the feature maps produced by ResNet need to be concatenated with a target speaker embedding. The concatenated features are then fed to the BiLSTM layers and fully connected layers. Since TS-VAD training needs large computing resources, we only train a model with $S = 2$ for statistical pooling. The number of target speakers embedding $N$ is set to 8. The parameters of front ResNet34 are initialized from our speaker embedding model. 

The model is first trained on the simulated LibriSpeech for 10 epochs with front ResNet34 frozen, and then it is trained for another 10 epochs with all parameters. Next, we transfer this model to VoxConverse data by training on the simulated DEV402 for 10 epochs. Finally, we fine-tune the model on DEV402 for 200 epochs and validate on VAL46. The learning rate is 0.0001 when training on simulated data and 0.00001 during the fine-tuning stage. The model is optimized by BCE loss and Adam optimizer. The input is 16s chunked wav, and the acoustic feature is 80-dim log Mel-filterbank energies with a frame length of 25ms and a frameshift of 10ms. 

\subsubsection{Inference}
For inference, the non-speech regions are first removed by VAD, and the wavs are split into 16s chunks. Next, speaker embeddings are extracted given the results from a clustering-based method. We only consider those speaker embeddings with 16s or longer speech. If the number of speakers is less than 8, we use zero-vectors as the fake embeddings. If it is greater than 8, we discard the speaker embeddings with shorter speech, but their labels are kept in the final results. The threshold for speaker decision is set to 0.8. 

\subsection{2-speaker TS-VAD for Overlap Detection}
The training data, model configuration, and training process are the same as the TS-VAD in Sec. \ref{sec:TSVAD} except that the number of target speaker $N$ is 2. For each recording, we select at most 5 speakers with the longest speech for inference, which provides at most ${5 \choose 2}=10$ pairs of target speaker embeddings. After obtaining the speaker decision of each pair of speakers by a threshold of 0.8, we update the results with the detected overlapped speech regions. This 2-speaker TS-VAD method can be applied to any data without considering the number of speakers. In this challenge, we only consider modifying the overlapped speech regions, but the single speaker region can also be iteratively refined according to the output from this 2-speaker TS-VAD model. And we will do more experiments later. 

\subsection{Data Augmentation}
We perform online data augmentation \cite{cai2020fly} with MUSAN and RIRs corpus. For background additive noise, we use ambient noise, music, television, and babble noise. For reverberation, we perform convolution with 40,000 simulated room impulse responses from small and medium rooms. The data augmentation is employed for all models which take acoustic features as input. 

\subsection{System Fusion}
To further improve the performance and robustness, we fuse our systems by DOVER-Lap \cite{doverlap}. 

\section{Experimental Results}

Table \ref{tab:result} shows the results on both VAL46 and the challenge test set. For the clustering-based system, the AHC method achieves a DER of $4.42\%$ on VAL46 and $6.77\%$ on the test set, which is our first submission. We employed our best VAD model for all systems on VAL46, but our first submission includes a poor VAD model, and it may not correctly reveal the improvements brought by OSD. 

For the 2nd submission, we fused systems 3, 4, and 7 using DOVER-Lap with custom weight tuned on VAL46, and we obtained a DER of $4.41\%$ on VAL46 and $5.36\%$ on the test set. 

For the 3rd submission, we fuse systems 3, 4, 5, and 7 using DOVER-Lap with custom weight tuned on VAL46. However, the DER on the test set goes lower. We did not know if it was system 5 that reduced the performance, so we directly submitted system 5, and it shows a DER of $5.45\%$. Therefore, the reason may be that we tuned the weights so aggressively, and the fused system is over-fitted on VAL46.

Finally, we fused systems 3, 4, 5, 6, and 8 with rank-based weighting and achieve a DER of $5.07\%$ on the test set, which ranked 1st at the VoxSRC 2021. 

\section{Conclusions}

In this report, we describe our system for the VoxSRC 2021. To achieve better performance, we mainly focus on the overlapped speech detection. We employ overlap detection and TS-VAD to reduce the missed speaker error. In addition, we propose a 2-speaker TS-VAD framework to detect the overlapped speech between each pair of speakers. Our experiments show that detecting the overlapped speech regions can significantly improve performance.

\bibliographystyle{IEEEtran}

\bibliography{mybib}

\begin{thebibliography}{10}
\providecommand{\url}[1]{#1}
\csname url@samestyle\endcsname
\providecommand{\newblock}{\relax}
\providecommand{\bibinfo}[2]{#2}
\providecommand{\BIBentrySTDinterwordspacing}{\spaceskip=0pt\relax}
\providecommand{\BIBentryALTinterwordstretchfactor}{4}
\providecommand{\BIBentryALTinterwordspacing}{\spaceskip=\fontdimen2\font plus
\BIBentryALTinterwordstretchfactor\fontdimen3\font minus
  \fontdimen4\font\relax}
\providecommand{\BIBforeignlanguage}[2]{{%
\expandafter\ifx\csname l@#1\endcsname\relax
\typeout{** WARNING: IEEEtran.bst: No hyphenation pattern has been}%
\typeout{** loaded for the language `#1'. Using the pattern for}%
\typeout{** the default language instead.}%
\else
\language=\csname l@#1\endcsname
\fi
#2}}
\providecommand{\BIBdecl}{\relax}
\BIBdecl

\bibitem{microsoft}
X.~Xiao, N.~Kanda, Z.~Chen, T.~Zhou, T.~Yoshioka, S.~Chen, Y.~Zhao, G.~Liu,
  Y.~Wu, J.~Wu \emph{et~al.}, ``Microsoft speaker diarization system for the
  voxceleb speaker recognition challenge 2020,'' in \emph{ICASSP}.\hskip 1em
  plus 0.5em minus 0.4em\relax IEEE, 2021, pp. 5824--5828.

\bibitem{overlap}
K.~Boakye, B.~Trueba-Hornero, O.~Vinyals, and G.~Friedland, ``Overlapped speech
  detection for improved speaker diarization in multiparty meetings,'' in
  \emph{ICASSP}.\hskip 1em plus 0.5em minus 0.4em\relax IEEE, 2008, pp.
  4353--4356.

\bibitem{eend}
Y.~Fujita, S.~Watanabe, S.~Horiguchi, Y.~Xue, and K.~Nagamatsu, ``End-to-end
  neural diarization: Reformulating speaker diarization as simple multi-label
  classification,'' \emph{arXiv preprint arXiv:2003.02966}, 2020.

\bibitem{tsvad}
I.~Medennikov, M.~Korenevsky, T.~Prisyach, Y.~Khokhlov, M.~Korenevskaya,
  I.~Sorokin, T.~Timofeeva, A.~Mitrofanov, A.~Andrusenko, I.~Podluzhny,
  A.~Laptev, and A.~Romanenko, ``{Target-Speaker Voice Activity Detection: A
  Novel Approach for Multi-Speaker Diarization in a Dinner Party Scenario},''
  in \emph{INTERSPEECH}, 2020, pp. 274--278.

\bibitem{AMI}
J.~Carletta, S.~Ashby, S.~Bourban, M.~Flynn, M.~Guillemot, T.~Hain, J.~Kadlec,
  V.~Karaiskos, W.~Kraaij, M.~Kronenthal \emph{et~al.}, ``The ami meeting
  corpus: A pre-announcement,'' in \emph{International workshop on machine
  learning for multimodal interaction}.\hskip 1em plus 0.5em minus 0.4em\relax
  Springer, 2005, pp. 28--39.

\bibitem{ICSI}
A.~Janin, D.~Baron, J.~Edwards, D.~Ellis, D.~Gelbart, N.~Morgan, B.~Peskin,
  T.~Pfau, E.~Shriberg, A.~Stolcke \emph{et~al.}, ``The icsi meeting corpus,''
  in \emph{ICASSP}, vol.~1.\hskip 1em plus 0.5em minus 0.4em\relax IEEE, 2003,
  pp. I--I.

\bibitem{aishell4}
Y.~Fu, L.~Cheng, S.~Lv, Y.~Jv, Y.~Kong, Z.~Chen, Y.~Hu, L.~Xie, J.~Wu, H.~Bu
  \emph{et~al.}, ``Aishell-4: An open source dataset for speech enhancement,
  separation, recognition and speaker diarization in conference scenario,''
  \emph{arXiv preprint arXiv:2104.03603}, 2021.

\bibitem{DIHARDII}
N.~Ryant, K.~Church, C.~Cieri, A.~Cristia, J.~Du, S.~Ganapathy, and
  M.~Liberman, ``The second dihard diarization challenge: Dataset, task, and
  baselines,'' \emph{arXiv preprint arXiv:1906.07839}, 2019.

\bibitem{DIHARDIII}
N.~Ryant, P.~Singh, V.~Krishnamohan, R.~Varma, K.~Church, C.~Cieri, J.~Du,
  S.~Ganapathy, and M.~Liberman, ``The third dihard diarization challenge,''
  \emph{arXiv preprint arXiv:2012.01477}, 2020.

\bibitem{Voxceleb}
A.~Nagrani, J.~S. Chung, W.~Xie, and A.~Zisserman, ``Voxceleb: Large-scale
  speaker verification in the wild,'' \emph{Computer Speech \& Language},
  vol.~60, p. 101027, 2020.

\bibitem{Librispeech}
V.~Panayotov, G.~Chen, D.~Povey, and S.~Khudanpur, ``Librispeech: an asr corpus
  based on public domain audio books,'' in \emph{ICASSP}.\hskip 1em plus 0.5em
  minus 0.4em\relax IEEE, 2015, pp. 5206--5210.

\bibitem{Voxconverse2020}
J.~S. Chung, J.~Huh, A.~Nagrani, T.~Afouras, and A.~Zisserman, ``{Spot the
  Conversation: Speaker Diarisation in the Wild},'' in \emph{INTERSPEECH},
  2020, pp. 299--303.

\bibitem{MUSAN}
D.~Snyder, G.~Chen, and D.~Povey, ``{MUSAN}: {A} {Music}, {Speech}, and {Noise}
  {Corpus},'' \emph{arXiv:1510.08484}, 2015.

\bibitem{RIRs}
T.~Ko, V.~Peddinti, D.~Povey, M.~L. Seltzer, and S.~Khudanpur, ``A study on
  data augmentation of reverberant speech for robust speech recognition,'' in
  \emph{ICASSP}.\hskip 1em plus 0.5em minus 0.4em\relax IEEE, 2017, pp.
  5220--5224.

\bibitem{resnet}
K.~He, X.~Zhang, S.~Ren, and J.~Sun, ``Deep residual learning for image
  recognition,'' in \emph{CVPR}, 2016, pp. 770--778.

\bibitem{bilstm}
M.~Schuster and K.~K. Paliwal, ``Bidirectional recurrent neural networks,''
  \emph{IEEE transactions on Signal Processing}, vol.~45, no.~11, pp.
  2673--2681, 1997.

\bibitem{arcface}
J.~Deng, J.~Guo, N.~Xue, and S.~Zafeiriou, ``Arcface: Additive angular margin
  loss for deep face recognition,'' in \emph{CVPR}, 2019.

\bibitem{ffsvc20}
X.~Qin, M.~Li, H.~Bu, R.~K. Das, W.~Rao, S.~Narayanan, and H.~Li, ``The ffsvc
  2020 evaluation plan,'' \emph{arXiv preprint arXiv:2002.00387}, 2020.

\bibitem{LinLSTM}
Q.~Lin, R.~Yin, M.~Li, H.~Bredin, and C.~Barras, ``{LSTM Based Similarity
  Measurement with Spectral Clustering for Speaker Diarization},'' in
  \emph{INTERSPEECH}, 2019, pp. 366--370.

\bibitem{DIHARDII-LSTM}
Q.~Lin, W.~Cai, L.~Yang, J.~Wang, J.~Zhang, and M.~Li, ``{DIHARD II is Still
  Hard: Experimental Results and Discussions from the DKU-LENOVO Team},'' in
  \emph{Odyssey}, 2020, pp. 102--109.

\bibitem{dihard3tsvad}
Y.~Wang, M.~He, S.~Niu, L.~Sun, T.~Gao, X.~Fang, J.~Pan, J.~Du, and C.-H. Lee,
  ``Ustc-nelslip system description for dihard-iii challenge,'' \emph{arXiv
  preprint arXiv:2103.10661}, 2021.

\bibitem{cai2020fly}
W.~Cai, J.~Chen, J.~Zhang, and M.~Li, ``On-the-fly data loader and
  utterance-level aggregation for speaker and language recognition,''
  \emph{IEEE/ACM Transactions on Audio, Speech, and Language Processing},
  vol.~28, pp. 1038--1051, 2020.

\bibitem{doverlap}
D.~Raj, L.~P. Garcia-Perera, Z.~Huang, S.~Watanabe, D.~Povey, A.~Stolcke, and
  S.~Khudanpur, ``Dover-lap: A method for combining overlap-aware diarization
  outputs,'' in \emph{SLT}.\hskip 1em plus 0.5em minus 0.4em\relax IEEE, 2021,
  pp. 881--888.

\end{thebibliography}


\end{document}